\documentclass[aps,prx,showpacs,superscriptaddress,twocolumn,longbibliography]{revtex4-2}

\usepackage{amsfonts}
\usepackage{subfigure}
\usepackage{amsmath}
\usepackage{txfonts}
\usepackage{amssymb}
\usepackage{amsbsy} 
\usepackage{epsfig}
\usepackage{graphicx}
\usepackage{epstopdf}
\usepackage{titlesec}
\usepackage{appendix}
\usepackage{color}

\usepackage{mathdots}

\def\be{\begin{equation}} \def\ee{\end{equation}}
\def\bea{\begin{eqnarray}} \def\eea{\end{eqnarray}}

\def\be{{\bf e}}

\newcommand{\ket}[1]{|#1\rangle}

\begin{document}

\title{Solvable BCS-Hubbard Liouvillians in arbitrary dimensions}

\author{Xu-Dong Dai}\altaffiliation{These authors contributed equally to this work.}
 \affiliation{ Institute for
Advanced Study, Tsinghua University, Beijing,  100084, China }

\author{Fei Song}\altaffiliation{These authors contributed equally to this work.}
 \affiliation{ Institute for
Advanced Study, Tsinghua University, Beijing,  100084, China }
 \affiliation{Kavli Institute for Theoretical Sciences, Chinese Academy of Sciences, 100190 Beijing, China }

\author{Zhong Wang} \email{ wangzhongemail@tsinghua.edu.cn }
\affiliation{ Institute for
Advanced Study, Tsinghua University, Beijing,  100084, China }


\begin{abstract}
We construct a solvable Lindblad model in arbitrary dimensions, in which the Liouvillian can be mapped to a BCS-Hubbard model featuring an imaginary Hubbard interaction. The Hilbert space of the system can be divided into multiple sectors, each characterized by an onsite invariant configuration. The model exhibits bistable steady states in all spatial dimensions, which is guaranteed by the fermion-number parity. Notably, the Liouvillian gap exhibits a Zeno transition, below which the Liouvillian gap is linear with respect to the dissipation.  We also uncover a generic dimension-dependent gap behavior: In one dimension, the gap originates from multiple sectors with spectral crossing; in higher dimensions, a single sector determines the gap.  
\end{abstract}

\maketitle

\section{INTRODUCTION}

The competition between quantum correlations and the couplings to the environment leads to diverse physical consequences in open quantum systems. Recently, both theoretical and experimental progress has been made in understanding and utilizing such competition. There are theoretical proposals considering open quantum systems as promising platforms for quantum-state engineering \cite{kraus2008preparation, diehl2010dissipation, diehl2011topology, kastoryano2011dissipative, reiter2016scalable} and quantum computation \cite{beige2000quantum, verstraete2009quantum, kliesch2011dissipative, kastoryano2013precisely}. Meanwhile, the rapid developments of experimental techniques open up avenues for exploring open many-body quantum systems \cite{barreiro2011open, barontini2013controlling, fitzpatrick2017observation}.

When an open quantum system is surrounded by a Markovian environment, its time evolution is generally governed by the quantum master equation \cite{lindblad1976generators, 10.1093/acprof:oso/9780199213900.001.0001}.  The generator of the Lindblad equation (i.e., the Liouvillian) is a linear operator acting on the density matrix. Liouvillians are often studied by perturbative expansions \cite{reiter2012effective, vznidarivc2015relaxation, li2016resummation, shishkov2020perturbation} and numerical tools \cite{daley2014quantum, cui2015variational, kshetrimayum2017simple, nagy2019variational, weimer2021simulation}. However, its dimension is the square of the dimension of Hilbert space, making many-body Liouvillians even less numerically tractable than Hamiltonians. Thus, there have been considerable efforts in solving many-body Liouvillians exactly, including diagonalizing the complete spectrum and extracting steady states \cite{prosen2008third, medvedyeva2016exact, de2021constructing, nakagawa2021exact, vznidarivc2010exact, prosen2011open, prosen2011exact, karevski2013exact, prosen2014exact, de2021constructing}. Constructing a solvable Liouvillian is challenging, and most progress has been restricted to one dimension.

Here, we construct a Liouvillian that can be exactly solved in arbitrary dimensions. This model is inspired by the correspondence between Liouvillians and non-Hermitian Hamiltonians. Specifically, we construct a spinless fermionic dissipative model consisting of nearest-neighbor hoppings, BCS pairings, and on-site dephasing noise, which can be mapped to a form akin to a BCS-Hubbard model \cite{chen2018exactly}. This non-Hermitian Hamiltonian commutes with extensive local operators and is therefore solvable, which is reminiscent of the Kitaev honeycomb model \cite{kitaev2006anyons}. Notably, the ``Hubbard interaction'' in our model is imaginary and the physical interpretation is entirely different. We exactly obtain two steady states of this dissipative model and analyze the Liouvillian gap. From the dissipation dependence of the Liouvillian gap, we unveil a universal transition in all dimensions.

\section{BCS-HUBBARD LIOUVILLIAN AND ITS DIMENSIONAL-INDEPENDENT SOLVABLE STRUCTURES}

\subsection{Model} 

We consider an open system whose density matrix $\rho$ follows the master equation
\begin{equation}
\frac{d\rho}{dt}=-{i}[H_0,\rho]+\sum_l \left(L_{l}\rho L^{\dagger}_{l}-\frac{1}{2}\{ L^{\dagger}_{l}L_{l},\rho \}\right). \label{Lindblad}
\end{equation}
The system is placed on a $d$-dimensional bipartite lattice that includes $A, B$ sublattices. The Hamiltonian
\begin{equation}
H_0=\sum_{\langle i,j\rangle, i\in A}(t_{ij}c^{\dagger}_{i}c_{j}+\Delta_{ij}c^{\dagger}_{i}c^{\dagger}_{j}+\rm H.c.),\label{H}
\end{equation}
describes spinless fermions with both symmetric hoppings $t_{ij}=t_{ji}$ and staggered BCS pairings $\Delta_{ij}=-\Delta_{ji}$, and $\langle i,j\rangle$ denotes a pair of nearest-neighbor sites. Meanwhile, the dephasing process of this open system is controlled by the dissipators $L_{l}=\sqrt{\gamma}c^{\dagger}_{l}c_{l}=\sqrt{\gamma}n_l$ where $l \in A, B$. 

\begin{figure}
\includegraphics[width=5 cm, height=5 cm]{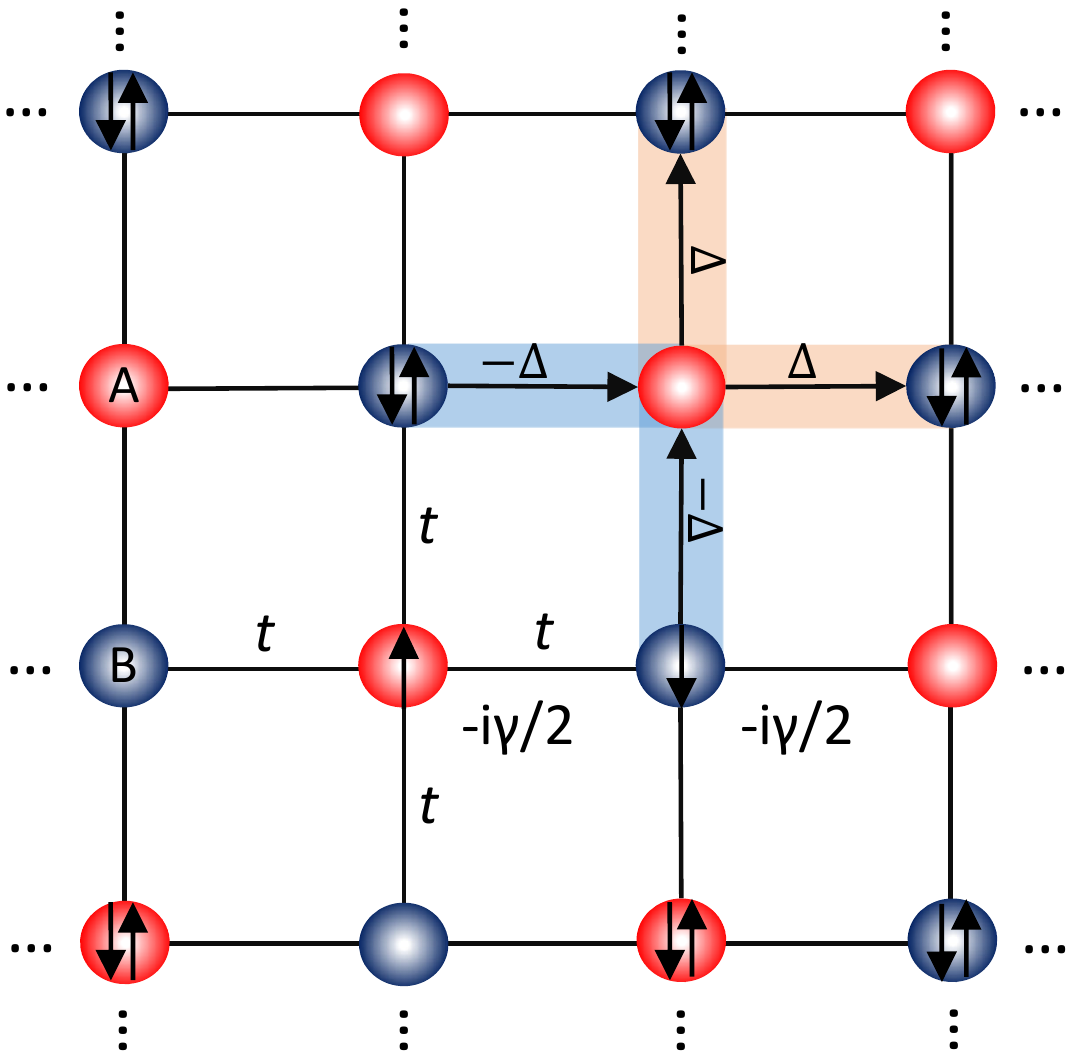}
\caption{Sketch of the non-Hermitian BCS-Hubbard Hamiltonian on a square lattice. We take hoppings $t_{ij}=t$ and BCS pairings $\Delta_{AB}=-\Delta_{BA}=\Delta$. The imaginary-Hubbard term $i\gamma(n_{l\uparrow}-\frac{1}{2})(n_{l\downarrow}-\frac{1}{2})-i\gamma/4$ is $-i\gamma/2$ when the site is occupied by a single fermion.}\label{BCSHubbard}
\end{figure}

The master equation can be compactly written as $d\rho/dt=\mathcal{L}[\rho]$ and the superoperator $\mathcal{L}$ is called Liouvillian (or Lindbladian). $\mathcal{L}$ can be mapped to a Hamiltonian-like operator by vectorizing the density matrix: $\rho=\sum_{nm}\rho_{nm} |n\rangle\langle m| \rightarrow |\rho\rangle=\sum_{nm}\rho_{nm} |n\rangle|m\rangle$. The fermionic operators acting on the density matrix by the left and right multiplication are then mapped to two sets of independent fermionic operators $c,c^\dagger$, and $\tilde{c}, \tilde{c}\dagger$ (see Appendix A and Refs. \cite{prosen2008third, dzhioev2012nonequilibrium}). The resultant expression reads 
\begin{equation}
    \begin{aligned}
\mathcal{L}=-{i}\sum_{\langle i,j\rangle, i\in A}[ t_{ij}(c^{\dagger}_{i}c_{j}-\tilde{c}^{\dagger}_{i}\tilde{c}_{j})+ \Delta_{ij}(c^{\dagger}_{i}c^{\dagger}_{j}+\tilde{c}^{\dagger}_{i}\tilde{c}^{\dagger}_{j})&\\
+{\rm H.c.}]+\gamma\sum_{l}(n_l-\frac{1}{2})(\tilde{n}_l-\frac{1}{2})-\frac{N\gamma}{4}&, \label{Lindbladian}
    \end{aligned}
\end{equation}
where $N$ denotes the total number of lattice sites. By further applying a transformation $H={i}U^{\dagger}\mathcal{L}U$ with the unitary matrix $U=\prod_{i\in A, j\in B}{\rm exp}[{{i}\pi/2(\tilde{c_{i}}^{\dagger}\tilde{c_{i}}-\tilde{c_{j}}^{\dagger}\tilde{c_{j}})}]$ and rewriting the fermion operators $c\to c_{\uparrow}$, $\tilde{c}\to c_{\downarrow}$, we transform the Liouvillian into a non-Hermitian BCS-Hubbard Hamiltonian  
\begin{equation}
   \begin{aligned}
H=\sum_{\langle i,j\rangle,i\in A,\sigma}(t_{ij}c^{\dagger}_{i\sigma}c_{j\sigma}+\Delta_{ij}c^{\dagger}_{i\sigma}c^{\dagger}_{j\sigma}+{\rm H.c.})&\\
+{i}\gamma \sum_{l\in A, B}(n_{l\uparrow}-\frac{1}{2})(n_{l\downarrow}-\frac{1}{2})-{i}\frac{N\gamma}{4}&\label{Hubbard}.
   \end{aligned} 
\end{equation}  
This form is reminiscent of the BCS-Hubbard model \cite{chen2018exactly}, but the 
Hubbard coupling $i\gamma$ is now imaginary and its physical meaning is completely different. An illustration in two dimensions is given in   Fig.~\ref{BCSHubbard}. Note that a complex Hubbard term can also be generated by two-body loss \cite{yamamoto2019theory, yamamoto2021collective, mazza2023dissipative}. 

\subsection{Solvable structures} 

For the sake of simplicity, we take $t_{ij}=t$ and $\Delta_{i\in A, j\in B}=\Delta$ to be translationally invariant. The more complex cases are discussed in Appendix B. To better reveal the  solvable structures of the Hamiltonian, we introduce two sets of Majorana fermions on A and B sublattices,
\begin{equation}
\begin{aligned}
A: c_{i\sigma}&=\frac{\alpha_{i\sigma}+{i}\beta_{i\sigma}}{2},\quad c^{\dagger}_{i\sigma}=\frac{\alpha_{i\sigma}-{i}\beta_{i\sigma}}{2};\nonumber\\
B: c_{j\sigma}&=\frac{\beta_{j\sigma}+{i}\alpha_{j\sigma}}{2},\quad  c^{\dagger}_{j\sigma}=\frac{\beta_{j\sigma}-{i}\alpha_{j\sigma}}{2}.\label{mbasis}
\end{aligned}
\end{equation}
In this Majorana representation, the Hamiltonian becomes
\begin{equation}
    \begin{aligned}
H=i\sum_{\langle i,j\rangle,i\in A,\sigma}(-\frac{t+\Delta}{2}\beta_{i\sigma}\beta_{j\sigma}+\frac{t-\Delta}{2}\alpha_{i\sigma}\alpha_{j\sigma})&\\
-{i}\frac{\gamma}{4}\sum_{l\in A,B}({i}\alpha_{l\uparrow}\alpha_{l\downarrow})({i}\beta_{l\uparrow}\beta_{l\downarrow})-{i}\frac{N\gamma}{4}&.
    \end{aligned}
\end{equation}
Importantly, the hopping term of $\alpha$-Majorana fermions vanishes when $t=\Delta$; then the quantities $D_{l}={i}\alpha_{l\uparrow}\alpha_{l\downarrow}$ are conserved on each site, i.e., $[D_{l},H]=0$ for all $l$. Using $\alpha_{l\sigma}^2=1$ and $\{\alpha_{l\uparrow}, \alpha_{l\downarrow}\}=0$, we know $D_l^2=1$. At the point $t=\Delta$, the Hilbert space is divided into $2^N$ different sectors marked by the conserved quantities $\{D_{l}=\pm1\}$, and the Hamiltonian in each sector reduces to
\begin{equation}
\tilde{H}(\{D_l\})=-{i}t\sum_{\langle i,j\rangle,i\in A,\sigma}\beta_{i\sigma}\beta_{j\sigma}-{i}\frac{\gamma}{4}\sum_{l\in A,B}D_{l}(i\beta_{l\uparrow}\beta_{l\downarrow})-{i}\frac{N\gamma}{4}.
\end{equation}
By combining two remaining Majorana fermions $\beta_{i\uparrow}$ and $\beta_{i\downarrow}$, we introduce new a-fermions as
\begin{equation}
    \begin{aligned}
&A: a_{i}=\frac{1}{2}(\beta_{i\uparrow}+{i}\beta_{i\downarrow}),\ a^{\dagger}_{i}=\frac{1}{2}(\beta_{i\uparrow}-{i}\beta_{i\downarrow});\\
&B: a_{j}=\frac{1}{2}(\beta_{j\downarrow}-{i}\beta_{j\uparrow}),\  a^{\dagger}_{j}=\frac{1}{2}(\beta_{j\downarrow}+{i}\beta_{j\uparrow}).\label{abasis}
    \end{aligned}
\end{equation}
Then, the above Hamiltonian transforms to
\begin{equation}
    \begin{aligned}
\tilde{H}(\{D_l\})=2t\sum_{\langle i,j \rangle, i\in A}(a^{\dagger}_{i}a_{j}+a^{\dagger}_{j}a_{i})-{i}\frac{\gamma}{2}\sum_{l\in A, B}D_{l}a^{\dagger}_{l}a_{l}&\\
+{i}\frac{\gamma}{4}\sum_{l\in A, B}(D_{l}-1)&.\label{afermion}
    \end{aligned}
\end{equation}
This Hamiltonian describes noninteracting spinless fermions living on a lattice with imaginary on-site potential. The two Hamiltonians $\tilde{H}(\{D_l\})$ and $\tilde{H}(\{-D_l\})$ are related by the charge-hole conjugation $a_{i\in \text{A}}\leftrightarrow a^\dagger_{i\in \text{A}},\ a_{j\in \text{B}}\leftrightarrow -a^\dagger_{j\in \text{B}} $, which ensures that two opposite sectors $\{D_{l}\}$ and $\{-D_{l}\}$ hold  the same spectrum. Since the first line of Eq.~(\ref{afermion}) is bilinear with fermion operators,  the Hamiltonian can be also written as
\begin{equation}
\tilde{H}(\{D_l\})=\sum_{i,j} h(\{D_{l}\})_{ij}a_i^\dagger a_j+{i}\frac{\gamma}{4}\sum_{l\in A, B}(D_{l}-1), \label{structure}
\end{equation}
where $h(\{D_{l}\})$ is an $N\times N$ matrix depending on the configuration of local conserved quantities $\{D_l\}$. This Hamiltonian, and therefore the original Liouvillian, can be solved by exactly diagonalizing $h(\{D_{l}\})$. Notably, this solvability is independent of the spatial dimension.  Coincidentally,  its 1d version can be mapped to the spin model discussed in Ref. \cite{shibata2019dissipative} via Jordan-Wigner transformation.

\section{EXACT SOLUTIONS FOR STEADY STATES AND LIOUVILLIAN GAP}

\subsection{Bistable steady states} 

The steady state which satisfies $\mathcal{L}[\rho_s]=0 $ can be mapped from the zero-energy state of $H$. We can exactly construct such states when $H$ becomes solvable at $t=\Delta$. From Eq.~(\ref{afermion}), it is straightforward to check that there are two possible zero-energy states, which are the vacuum state in the sector with all $D_{l}=+1$ and the fully occupied state in the sector with all $D_l=-1$. We write these two states as
\begin{equation}
|s_{U}^{+}\rangle=\prod_{l \in A, B}|0_a\rangle_{D_l=+1},\quad
|s_{U}^{-}\rangle=\prod_{l \in A, B}|1_a\rangle_{D_l=-1}.
\label{zerostate}
\end{equation}
These concise expressions have been obtained after a series of transformations. To recover the steady states in matrix form, we need inverse procedures. First, reexpress $|0_a\rangle_{D_l=+1}$ and $|1_a\rangle_{D_l=-1}$ in the basis $\{|n_{l\uparrow},n_{l\downarrow}\rangle\}$ as $|0_a\rangle_{D_l=+1}=(|00\rangle\pm{i}|11\rangle)/\sqrt{2}$ and $|1_a\rangle_{D_l=-1}=(|00\rangle\mp{i}|11\rangle)/\sqrt{2}$, where the signs depend on whether $l$ belongs to $A$ or $B$. Then,  converting $\{|n_{l\uparrow},n_{l\downarrow}\rangle\}$ to $\{|n_{l},\tilde{n}_{l}\rangle\}$  by the unitary matrix $U$, we find that the two zero modes of $\mathcal{L}$ are $|s^\pm\rangle=U|s_{U}^{\pm}\rangle$. At last, map the states $|s^\pm\rangle$ back to two matrices,
\begin{equation}
\begin{aligned}
&|s^{+}\rangle \to \rho^{+}=\prod_{l \in A, B}\frac{1}{2}(|0\rangle \langle0|_{l}-|1\rangle \langle1|_{l});\\
&|s^{-}\rangle \to \rho^{-}=\prod_{l \in A, B}\frac{1}{2}(|0\rangle \langle0|_{l}+|1\rangle \langle1|_{l})\label{steadystate}.
\end{aligned}
\end{equation}
Apparently,  one can easily verify that $\rho^-=I/2^N$ is a steady-state solution from Eq.~(\ref{Lindblad}) since all dephasing Lindblad operators are Hermitian. Moreover, although $\rho^+$ itself  is not a physical density matrix because of $\rm Tr \left(\rho^+ \right)=0$,  the linear combination of $\rho^{+}$ and $\rho^{-}$
\begin{equation}
\rho_{q}=\rho^{-}+q\rho^{+},\quad q\in [-1,1],
\end{equation}
contributes another steady state satisfying both $\mathcal{L}[\rho_q]=0 $ and $\rm Tr(\rho_q)=1$. The restriction of the parameter $q$  guarantees that the eigenvalues of $\rho_q$ can be interpreted as physical probabilities.  In particular, the two special combinations $\rho_e=\rho^{-}+\rho^{+}$ and $\rho_o=\rho^{-}-\rho^{+}$ have clear physical meanings. They correspond to maximally mixed states in the Hilbert space with even and odd particle number.  The system with two independent steady states which are $\rho_{e}$ and $\rho_{o}$ here is called bistable \cite{letscher2017bistability}. This is caused by the BCS pairing term in our model. The pairing term can only create and annihilate particles in pairs, so that the parity of particle number is conserved. More explicitly, if we define a fermion parity operator $S=\prod_{l \in A, B} (-1)^{n_l}$, the expectation value $\rm Tr \left(S\rho\right)$ is unchanged under the time evolution of Eq.~(\ref{Lindblad}). In other words, we have $\rm Tr (S\mathcal{L}[\rho])=\rm Tr\ (\mathcal{L}^\dagger[S]\rho)=0$, where $\mathcal{L}^\dagger[S]=i[H_0,S]+\sum_l(L_l^\dagger SL_l-\frac{1}{2}\{L_l^\dagger L_l,S\})$. With $[H_0, S]=0$ and $L_l=L_l^\dagger$, we can prove that $\mathcal{L}^{\dagger}[S]=0$ and $S/2^N$ is exactly the matrix $\rho^+$ that we have found.  Therefore, from the consideration of symmetry,  $\rho_e$ and $\rho_o$ are always the two steady states of the Liouvillian in Eq.~(\ref{Lindbladian}), regardless of whether the model is at the solvable point $t=\Delta$ or not. While the steady states do not contain much structure, the full spectrum enjoys richer features. 

\subsection{Liouvillian gap}  

We now investigate the Liouvillian gap at the solvable point $t=\Delta$.  The Liouvillian gap measures how fast an open system approaches its steady states, and its standard definition is
\begin{equation}
\Lambda=-\max_{m, {\rm Re} \left( \lambda_m\right)\neq 0} {\rm Re} \left( \lambda_m\right),\label{gap}
\end{equation}
where $ \{\lambda_m\}$ are the eigenvalues of Liouvillian. From the solvable structure in Eq.~(\ref{structure}), we know that for this model, $\lambda_m$ can be constructed by the single-particle eigenenergies $E_\alpha$ of  $h(\{D_l\})$, like
\begin{equation}
\lambda_m=-i\sum_{\alpha} m_\alpha E_\alpha+\frac{\gamma}{4}\sum_l (D_l-1),\label{lambda-E}
\end{equation}
where $m_\alpha=0,1$ denotes the occupation number of single-particle states. Solving $E_\alpha$ for a given $\{D_l\}$ is relatively easy, while searching the slowest-decay mode whose eigenvalue $\lambda_m$ has the maximal nonzero real part from exponentially many configurations is still cumbersome. However, we will theoretically deduce and numerically verify that only a very few configurations are important.

Hereafter, we will denote the configuration as an $n$-flipped configuration when there are $n$ sites giving $D_l=-1$. Since the spectrums of configurations $\{D_{l}\}$ and $\{-D_{l}\}$ are identical, we only have to consider the configurations where $D_l=-1$ on at most half of sites, namely $0\leq n\leq N/2$.  As a warm-up,  we first examine the $0$-flipped one with $D_l=1$ on all sites. $h(\{D_l=1\})$ maintains translational invariance under periodic boundary conditions (PBCs),  and one can diagonalize its eigenenergies as $E({\bf{k}})=4t\sum_{\alpha=x,y,\ldots}\cos \left(k_\alpha\right)-{i}\gamma/2$ by Fourier transformation. For this configuration, the Liouvillian eigenvalues are $\lambda_m=-i\sum_{\bf{k}} m_{\bf{k}} E(\bf{k})$ with $m_{\bf{k}}=0,1$. Consistent with Eq.~(\ref{zerostate}), the vacuum state gives $\lambda_m=0$. The maximal real part of all other nonzero $\lambda_m$'s is $-\gamma/2$. For the 1-flipped configuration, it is straightforward to check that the vacuum state also gives $\lambda_{m}=-\gamma/2$. To obtain the Liouvillian gap, we need to compare $-\gamma/2$ with the maximal real part of the Liouvillian eigenvalues from other configurations. 

\begin{figure}
\includegraphics[width=4.2cm, height=4.2cm]{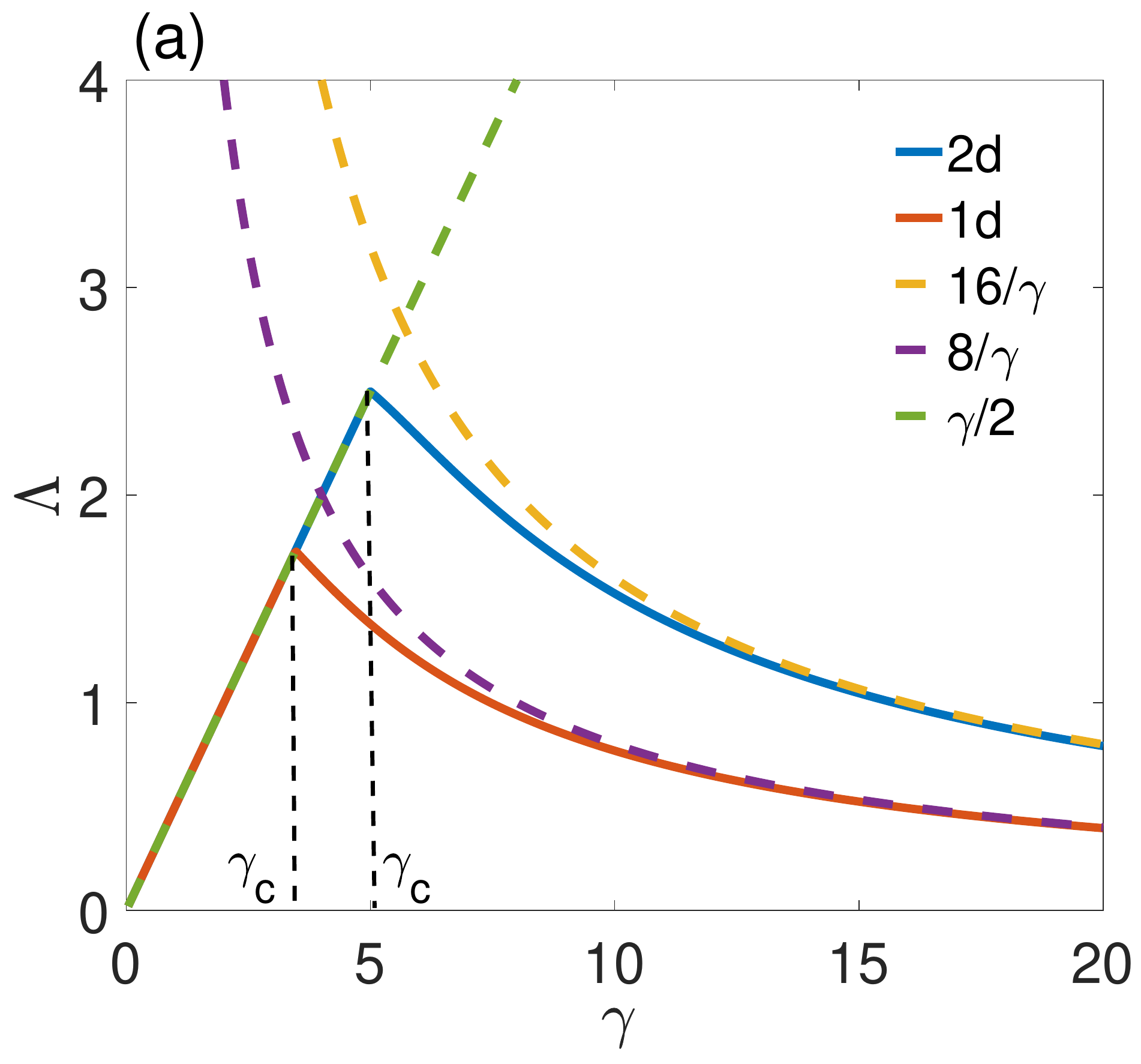}
\includegraphics[width=4.2cm, height=4.2cm]{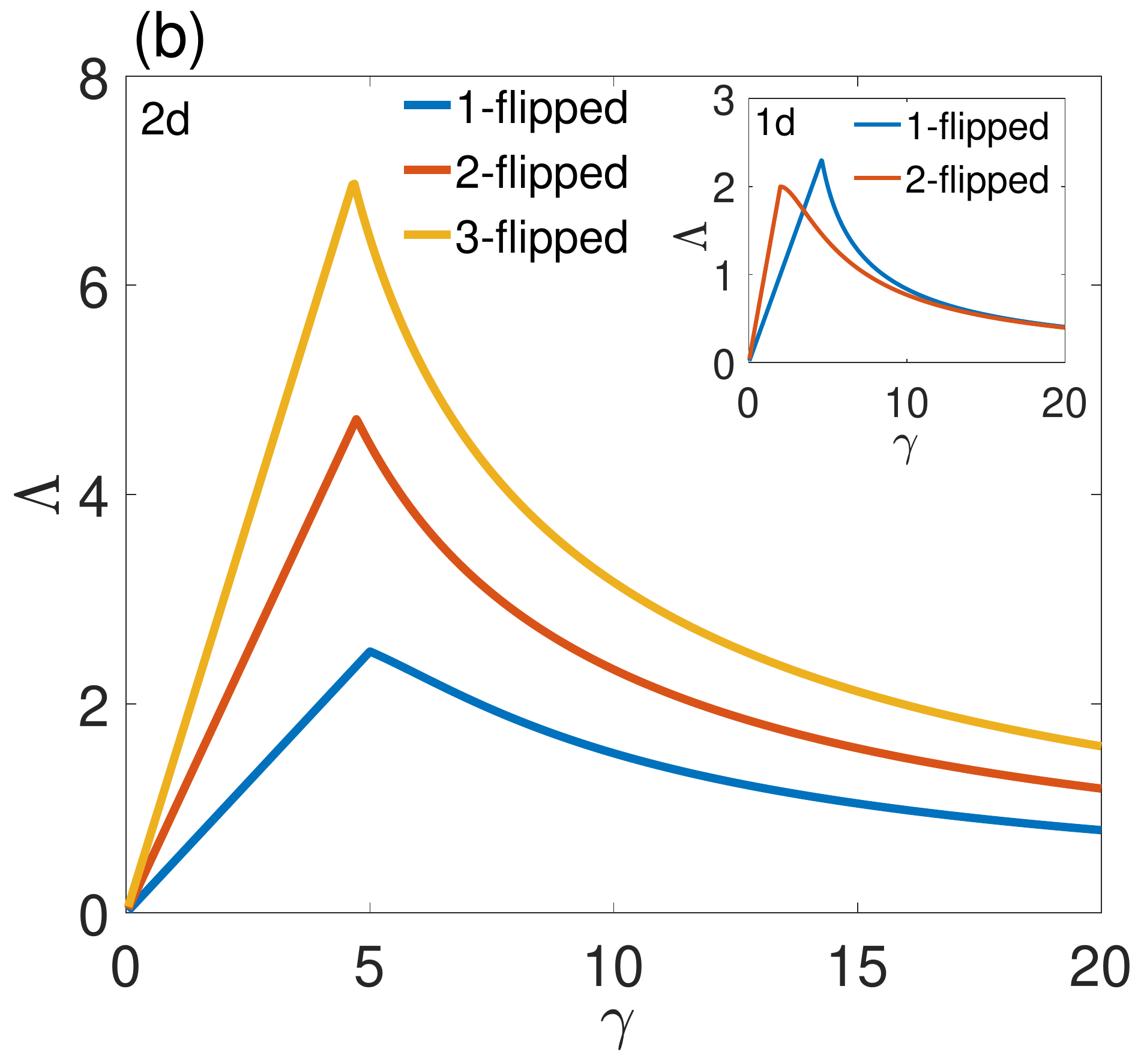}
\caption{(a)  The $\gamma$ dependence of the Liouvillian gap in 1d (red) and 2d (blue). The dashed lines represent the theoretical predictions. (b) compares the Liouvillian eigenvalues from several typical configurations. Each line depicts the $\gamma$ dependence of the absolute value of the maximal real part for one configuration. We select the $n$-flipped configuration where $D_l=-1$ on a line segment of length $n$. The solid lines in (a) are unions of the lowest part of the lines in (b). All eigenvalues are obtained from diagonalizing $h(\{D_l\})$ under PBCs. $N=40$ in 1d; $N=40\times40$ in 2d. $t=\Delta=1$. The inset in (b) gives the 1d situation and spectral crossing of sectors occurs, while in the main figure of (b), this will not happen in 2d. } \label{fig:gap}
\end{figure}

When dissipation is very weak (i.e., $\gamma\ll t$), we find that ${\rm Re} (\lambda_m)\leq-\gamma/2$ is satisfied by all nonzero Liouvillian eigenvalues. Consequently,  the Liouvillian gap follows
\begin{equation}
\Lambda(\gamma\ll t)=\gamma/2.
\end{equation}
[See Fig.~\ref{fig:gap}(a)].
According to Eq.~(\ref{lambda-E}), the maximal real part of the Liouvillian eigenvalues from one $n$-flipped configuration is $M_n=\sum_{\alpha, {\rm Im}(E_\alpha)>0} {\rm Im} (E_\alpha)-n\gamma/2$. Exploiting the fact that the eigenstates of $h(\{D_l\})$ are all extended when $\gamma=0$, to the first order of $\gamma$, ${\rm Im}(E_\alpha)\approx-\gamma/2+\gamma O(n/N)$ \footnote{Here we take $\delta h=-i\frac{\gamma}{2}\sum_{l}D_{l}a^{\dagger}_{l}a_{l}$ as the perturbation to $h(\{D_{l}\})$. The first-order result can be given by the standard perturbation theory}. It is obvious that ${\rm Im}(E_\alpha)$ are all negative when $n\ll N$. Thus, we have $M_n=-n\gamma/2$ when only a few $D_{l}$ are flipped. When $n/N$ is finite, the imaginary potential in Eq.~(\ref{structure}) acts like some sort of on-site disorder that may make $a$ fermions localize. This localization will reduce the imaginary energy cost $n\gamma/2$ from flipping $D_l$'s. However, we should notice that a sufficiently small $\gamma$ can only induce a weak localization. It implies that even though $M_n=-n\gamma/2$ no longer holds, the $M_n$ with a finite $n/N$ and a small $\gamma$ is still at the order of $\gamma O(n)$. Thus, we conclude that $M_1=-\gamma/2$ is indeed the upper bound of all nonzero ${\rm Re} (\lambda_m)$ in the small-$\gamma$ limit. The slowest-decay mode with ${\rm Re} (\lambda_m)=-\gamma/2$  resides in both $1$-flipped and $0$-flipped  configurations.

On the other hand,  when $\gamma\gg t$, the dissipative process becomes prominent. In this limit,  the imaginary Hubbard interaction ${i}\gamma \sum_{l}(n_{l\uparrow}-\frac{1}{2})(n_{l\downarrow}-\frac{1}{2})$ and the constant $-i\gamma N/4$ dominate in Eq.~(\ref{Hubbard}). When each site is either double occupied (as $n_{l\uparrow}=n_{l\downarrow}=1$) or empty (as $n_{l\uparrow}=n_{l\downarrow}=0$), these two contributions almost cancel each other, resulting in states with nearly zero energies.  These states form a subspace that is invariant under the action of the projection operator $P=\prod_l (1-n_{l\uparrow}-n_{l\downarrow}+2n_{l\uparrow}n_{l\downarrow})$. In this subspace, the remaining kinetic term $H_K$ of Eq.~(\ref{Hubbard}) generates an effective Hamiltonian whose leading order is formally $H_\text{eff}=-iPH_K^2P/\gamma$.  Substituting $H_K=\sum_{\langle i,j\rangle,i\in A,\sigma}(t c^{\dagger}_{i\sigma}c_{j\sigma}+\Delta c^{\dagger}_{i\sigma}c^{\dagger}_{j\sigma}+{\rm h.c.})$, we can derive
\begin{equation}
\begin{aligned}
H_{\text{eff}}=-\sum_{\substack{\langle i,j\rangle, i\in A;\\\sigma, \sigma'}}\frac{it^2}{\gamma}P\left[(c^{\dagger}_{i\sigma}c_{j\sigma}+c^{\dagger}_{j\sigma}c_{i\sigma})(c^{\dagger}_{i\sigma'}c_{j\sigma'}+c^{\dagger}_{j\sigma'}c_{i\sigma'})\right]P&\\ \sum_{\substack{\langle i,j\rangle, i\in A;\\\sigma, \sigma'}}\frac{i\Delta^2}{\gamma}P\left[(c^{\dagger}_{i\sigma}c^{\dagger}_{j\sigma}+c_{j\sigma}c_{i\sigma})(c^{\dagger}_{i\sigma'}c^{\dagger}_{j\sigma'}+c_{j\sigma'}c_{i\sigma'})\right]P&.
\end{aligned}
\label{effective1}
\end{equation}
The first and the second line in Eq.~(\ref{effective1}) account for second-order perturbations via hoppings and pairings, respectively. Considering that the subspace is locally two-dimensional, we further define
\begin{equation}
\begin{aligned}
\tau_l^x&:=P(c^{\dagger}_{l\uparrow}c^{\dagger}_{l\downarrow}+c_{l\downarrow}c_{l\uparrow})P, \\
\tau_l^y&:=iP(c^{\dagger}_{l\uparrow}c^{\dagger}_{l\downarrow}-c_{l\downarrow}c_{l\uparrow})P,\\
\tau_l^z :=1-&2Pc^{\dagger}_{l\uparrow}c_{l\uparrow}P=1-2Pc^{\dagger}_{l\downarrow}c_{l\downarrow}P,
\end{aligned}\label{tau}
\end{equation}
as the Pauli matrices with respect to the basis $|n_{l\uparrow}=0, n_{l\uparrow}=0\rangle={|\uparrow\rangle}=(1,0)^{\rm T}$ and $|n_{l\uparrow}=1, n_{l\downarrow}=1\rangle=|\downarrow\rangle=(0,1)^{\rm T}$. In this spin representation, Eq.~(\ref{effective1}) turns to
\begin{equation}
H_\text{eff}=-i\sum_{\langle i,j \rangle, i\in A}[J_{\perp}(-\tau^{z}_{i}\cdot \tau^{z}_{j}+\tau^{x}_{i}\cdot \tau^{x}_{j})+J(\tau^{y}_{i}\cdot \tau^{y}_{j}+1)],\label{AFM}
\end{equation}
where $J_{\perp}=(t^{2}-\Delta^{2})/\gamma$, $J=(t^{2}+\Delta^{2})/\gamma$, and $0<|J_{\perp}|\leq J$. At the solvable point $t=\Delta$, the effective Hamiltonian is simplified to an Ising model without quantum fluctuations in arbitrary dimensions:
\begin{equation}
H_{\text{eff}}=-i\frac{2t^2}{\gamma}\sum_{\langle i,j \rangle, i\in A}(\tau^{y}_{i}\cdot \tau^{y}_{j}+1),\label{Ising}
\end{equation}
which manifests the solvability of the original model. For this Ising model, there are two spin patterns giving zero energies: $\tau_{\forall i\in A}^y=1, \tau_{\forall j\in B}^y=-1$ and $\tau_{\forall i\in A}^y=-1, \tau_{\forall j\in B}^y=1$, which can be translated to the states $|s_{U}^{\pm}\rangle$  found in Eq.~(\ref {zerostate}) via the relation
\begin{equation}
\begin{aligned}
&A: \tau^y_{i} \ket{0_a}_{D_i=+1}=\ket{0_a}_{D_i=+1},\ \tau^y_{i} \ket{1_a}_{D_i=-1}=-\ket{1_a}_{D_i=-1} ;\\
&B:  \tau^y_{j} \ket{1_a}_{D_j=-1}=\ket{1_a}_{D_j=-1},\ \tau^y_{j} \ket{0_a}_{D_j=+1}=-\ket{0_a}_{D_j=+1}. \label{tau-D}
\end{aligned}
\end{equation}
Accordingly, the two zero modes of $H_\text{eff}$, just like $|s_{U}^{\pm}\rangle$, can produce the aforementioned bistable steady states. In addition to the steady states, the above relation together with $H_\text{eff}$ will help us acquire more information about the large-$\gamma$ limit. From Eq. (\ref{tau-D}), we find that flipping $\tau_l^y$ is equivalent to flipping $D_l$ and simultaneously creating (or annihilating) an $a$-fermion on the site $l$.  A one-to-one correspondence between Ising variables ($\tau_l^y$) and conserved charges ($D_l$) can be built, under which $D_{i\in A}=\tau_{i\in A}^y, D_{j\in B}=-\tau_{i\in B}^y$, and  $H_\text{eff}=i2t^2/\gamma\sum_{\langle i,j \rangle, i\in A}(D_i D_j-1)$. Generally, a configuration $\{D_l\}$ splits into domains with only $D_l=1$ or $D_l=-1$, and the formation of the domain walls costs nonzero energy. When the domain walls of $\{D_l\}$ intersect $L_D$ links, the energy equals to $-i4t^2L_D/\gamma$. By multiplying $-i$, the energies of $H_\text{eff}$ are mapped to the low-lying Liouvillian eigenvalues near zero, and therefore the Liouvillian gap in the large-$\gamma$ limit is simply
\begin{equation}
\Lambda (\gamma\gg t)=\frac{4t^2}{\gamma} \min_{L_D\neq0} L_D =\frac{8t^2d}{\gamma},
\end{equation}
where we have used $\min_{L_D\neq0} L_D=2d$ under PBCs in $d$ dimensions. Obviously, in the large-$\gamma$ limit,  the slowest-decay mode must belong to the configurations with $L_D=2d$. In two and higher dimensions, only $1$-flipped [and $(N-1)$-flipped] configurations are eligible. The 1-dimensional case is more complex;  $L_D=2$ is satisfied in all configurations containing only one connected domain with $D_l=-1$. To find which of them contributes the slowest-decay mode,  higher-order corrections to $H_\text{eff}$ are needed. However, even in one dimension, at most $N$ configurations are worth considering;  numerically extracting the exact Liouvillian gap from these configurations is only polynomial-hard. Notably, we find that the Liouvillian gap originates from two sectors in 1d, while in two and higher dimensions, the gap is solely determined by a single sector. In 1d, the slowest-decay mode contributing to the Liouvillian gap moves from the 1-flipped to the 2-flipped sector with the increasing of $\gamma$. See Fig.~\ref{fig:gap}(b) for an illustration. 

We have shown that the Liouvillian gap is proportional to a weak $\gamma$ while is reversely proportional to a strong $\gamma$. Naturally, a transition between these two qualitatively distinct behaviors is anticipated. This can be called a ``Zeno transition'' because freezing quantum dynamics (i.e., relaxation time approaching infinity) by increasing dissipation is analogous to the quantum Zeno effect \cite{vasiloiu2018enhancing, nakagawa2021exact}. The Zeno transition from $\Lambda (\gamma\ll t)=\gamma/2$ to $\Lambda (\gamma\gg t)=8t^2d/\gamma$ is a universal feature of this solvable model in all dimensions. We can estimate its critical point by $\gamma_c\sim4t\sqrt{d}$ from $\gamma_c/2 \sim8t^2d/\gamma_c$. Physically, the Zeno transition originates from the suppression of all possible single-fermion occupations by strong dissipation, which leads to a separation in the decay rates of the states with and without single-fermion occupations and thus permits the aforementioned second-order perturbation analysis. This is a general  mechanism that does not rely on exact solvability. Thus, we also expect that the Zeno transition similarly happens for $t\neq\Delta$. 


To confirm this picture, we illustrate numerical results for the Zeno transitions in both 1d and 2d [Fig.~\ref{fig:gap}(a)]. The small-$\gamma$ and large-$\gamma$ limits of the Liouvillian gap perfectly match our theoretical predictions. We also check which configuration is associated with the Liouvillian gap in Fig.~\ref{fig:gap}(b). Only the $1$-flipped configuration matters in 2d and higher dimensions, while a switch between $1$-flipped and $2$-flipped configurations is found in 1d. Therefore, the spectrum crossing of different sectors which contributes to the Liouvillian gap occurs exclusively in 1d. Additionally, we examine the wavefunction of the slowest-decay mode in the limit of large $\gamma$ and observe its localization at the site where $D_l$ undergoes a flip. This localization can be interpreted as a bound state between the flipped $D_l$ and the $a$ fermions. The Zeno transition corresponds to the formation of this bound state.


\section{Conclusions} 

We construct a solvable Liouvillian in arbitrary dimensions, where the dimension-independent solvability is facilitated by the presence of appropriate BCS pairings in the Hamiltonian. In all dimensions, we find bistable steady states and the Zeno transition of the Liouvillian gap. Notably, these phenomena persist even when the model deviates from the solvable regime. Quite a few aspects of this solvable Liouvillian remain to be explored. For example, it is interesting to investigate the intrinsic non-Hermitian degeneracies (i.e., exceptional points \cite{heiss2004exceptional}) of this Liouvillian and their physical consequences. Previous studies on similar topics are mostly on a few qubits \cite{khandelwal2021signatures, chen2022decoherence, minganti2019quantum}, whereas the solvable structure here enables investigating a many-body system. Thus, our solvable Liouvillian could offer a benchmarking model for theories of open quantum systems in dimensions greater than 1.


\section*{ACKNOWLEDGMENTS}

This work is supported by NSFC under Grant No. 12125405.

\appendix

\section*{APPENDIX A: MAPPING OF FERMIONIC OPERATORS}
When mapping the fermionic operators, we have to pay attention to the sign:
\begin{equation}
\begin{aligned}
c^{\dagger}&|m\rangle\langle n| \to c^{\dagger}|mn\rangle,\\
c&|m\rangle\langle n| \to c|mn\rangle,\\
&|m\rangle\langle n|c^{\dagger} \to (-1)^{N_{m}+N_{n}-1}\tilde{c}|mn\rangle,\\
&|m\rangle\langle n|c \to (-1)^{N_{m}+N_{n}}\tilde{c}^{\dagger}|mn\rangle,
\end{aligned}
\end{equation}
where $N_{m}$ is the particle number of state $|m\rangle$ and the factor is to keep anticommutation relations of these two sets of independent fermions $\{c_{i},c^{\dagger}_{j}\}=\{\tilde{c}_{i},\tilde{c}^{\dagger}_{j}\}=\delta_{ij}$, $\{c_{i},c_{j}\}=\{\tilde{c}^{\dagger}_{i},\tilde{c}^{\dagger}_{j}\}=0$ and $\{c_i,\tilde{c}_{j}\}=\{c_i^{\dagger},\tilde{c}_{j}^{\dagger}\}=\{c_i,\tilde{c}_{j}^{\dagger}\}=\{c_i^{\dagger},\tilde{c}_{j}\}=0$.

\section*{APPENDIX B: MORE GENERAL SOLVABLE CASES}
In any dimensional lattice, the bonds between $A$ and $B$ sublattices can be divided into two classes $A\rightarrow B$ and $B\rightarrow A$ in the positive direction. Now, we define $t_{i,\bf a}=t_{AB}$, $\Delta_{i,\bf a}=\Delta_{AB}$ and $t_{i,-\bf a}=t_{BA}$, $\Delta_{i,-\bf a}=-\Delta_{BA}$, where $i\in A$ and $\bf a$ is the nearest neighbor-vector in the positive direction. Then the $p$-wave BCS-Hubbard model is
\begin{equation}
\begin{aligned}
H=\sum_{i\in A ,\bf a,\sigma}&(t_{i,\bf a}c^{\dagger }_{i\sigma}c_{i+\bf a\sigma}+\Delta_{i,\bf a}c^{\dagger}_{i\sigma}c^{\dagger}_{i+\bf a\sigma}+t_{i,\bf -a}c^{\dagger}_{i-\bf a\sigma}c_{i\sigma}-\Delta_{i,-\bf a}c^{\dagger}_{i-\bf a\sigma}c^{\dagger}_{i\sigma}\\
&+{\rm H.c.})
+{i}\gamma \sum_{l}(n_{l\uparrow}-\frac{1}{2})(n_{l\downarrow}-\frac{1}{2})-{i}\frac{N\gamma}{4},
\end{aligned}
\end{equation}
where $n_{l\sigma}=c^{\dagger}_{l\sigma}c_{l\sigma}$, $N$ is the number of sites, and $i\gamma$ is the imaginary Hubbard interaction. With the two sets of Majorana fermions on A and B sublattices,
\begin{equation}
\begin{aligned}
&A: c_{i\sigma}=\frac{\alpha_{i\sigma}+{i}\beta_{i\sigma}}{2},\ c^{\dagger}_{i\sigma}=\frac{\alpha_{i\sigma}-{i}\beta_{i\sigma}}{2};\\ 
&B: c_{j\sigma}=\frac{\beta_{j\sigma}+{i}\alpha_{j\sigma}}{2},\  c^{\dagger}_{j\sigma}=\frac{\beta_{j\sigma}-{i}\alpha_{j\sigma}}{2},
\end{aligned}
\label{mbasis}
\end{equation}
the Hamiltonian  can be rewritten in the Majorana fermion basis:
\begin{equation}
\begin{split}
H=i\sum_{i\in A,\bf a,\sigma}(\frac{t_{i,\bf a}-\Delta_{i,\bf a}}{2}\alpha_{i\sigma}\alpha_{i+\bf a\sigma}+\frac{t_{i,-\bf a}-\Delta_{i,-\bf a}}{2}\alpha_{i\sigma}\alpha_{i-\bf a\sigma})&\\
-(\frac{t_{i,-\bf a}-\Delta_{i,-\bf a}}{2}\beta_{i\sigma}\beta_{i-\bf a\sigma}+\frac{t_{i,\bf a}+\Delta_{i,\bf a}}{2}\beta_{i\sigma}\beta_{i+\bf a \sigma})&\\
-{i}\frac{\gamma}{4}\sum_{l}({i}\alpha_{l\uparrow}\alpha_{i\downarrow})({i}\beta_{l\uparrow}\beta_{i\downarrow})-{i}\frac{N\gamma}{4}&.
\end{split}
\end{equation}
When $t_{i,\bf a}=\Delta_{i,\bf a}$ and $t_{i,-\bf a}=\Delta_{i,-\bf a}$, we have the conserved on-site quantity $D_{l}=i\alpha_{l\uparrow}\alpha_{l\downarrow}$ $(D^{2}_{l}=1)$ and the system becomes noninteracting.  Introduce a set of new $a$ fermions:
\begin{equation}
\begin{aligned}
&A: a_{i}=\frac{1}{2}(\beta_{i\uparrow}+{i}\beta_{i\downarrow}),\ a^{\dagger}_{i}=\frac{1}{2}(\beta_{i\uparrow}-{i}\beta_{i\downarrow});\\
&B: a_{j}=\frac{1}{2}(\beta_{j\downarrow}-{i}\beta_{j\uparrow}),\  a^{\dagger}_{j}=\frac{1}{2}(\beta_{j\downarrow}+{i}\beta_{j\uparrow}).
\end{aligned}
\label{abasis}
\end{equation}
The noninteracting Hamiltonian in the $a$-fermion basis is
\begin{equation}
\begin{aligned}
H=\sum_{i\in A,\bf a}2t_{i,\bf a}(a^{\dagger}_{i}a_{i+\bf a}+a^{\dagger}_{i+\bf a}a_{i})+2t_{i,-\bf a}(a^{\dagger}_{i}a_{i-\bf a}+a^{\dagger}_{i-\bf a}a_{i})&\\
-{i}\frac{\gamma}{2}\sum_{l}D_{l}a^{\dagger}_{l}a_{l}+{i}\frac{\gamma}{4}\sum_{l}(D_{l}-1)&.
\end{aligned}
\end{equation}
The discussion about the steady states and Liouvillian gap can be simplified in each sector with fixed $D_{l}$.

\section*{APPENDIX C: DETAILS IN CONSTRUCTING STEADY-STATE SOLUTIONS}
There is a four-dimensional Hilbert space expanded by the basis $|n_{\uparrow},n_{\downarrow}\rangle$ on each site. We have $D_{A}={i}\alpha_{A\uparrow}\alpha_{A\downarrow}={i}(c^{\dagger}_{A\uparrow}+c_{A\uparrow})(c^{\dagger}_{A\downarrow}+c_{A\downarrow})$. Choose the order of bases to be $\{|0_{\uparrow}0_{\downarrow}\rangle_{A},|0_{\uparrow}1_{\downarrow}\rangle_{A},|1_{\uparrow}0_{\downarrow}\rangle_{A},|1_{\uparrow}1_{\downarrow}\rangle_{A}\}$ and it is easy to get the matrix form of $D_{A}$ under the $|n_{\uparrow},n_{\downarrow}\rangle_{A}$ basis,
\begin{equation}
D_{A}=
\begin{pmatrix}
0&0&0&-{i} \\0&0&-{i}&0\\0&{i}&0&0 \\{i}&0&0 &0
\end{pmatrix}.
\end{equation}
The eigenvalues and the corresponding eigenvectors are
\begin{equation}
\begin{aligned}
+1 \quad |\tilde{0}\rangle_{A}&=\frac{1}{\sqrt{2}}(|00\rangle_{A}+{i}|11\rangle_{A}),\quad |\tilde{1}\rangle_{A}=\frac{1}{\sqrt{2}}(|01\rangle_{A}+{i}|10\rangle_{A});\nonumber\\
-1 \quad
|\tilde{0}\rangle_{A}&=\frac{1}{\sqrt{2}}({i}|10\rangle_{A}-|01\rangle_{A}),\quad|\tilde{1}\rangle_{A}=\frac{1}{\sqrt{2}}(|00\rangle_{A}-{i}|11\rangle_{A}).
\end{aligned}
\end{equation}
The same analysis of $D_{B}$:
\begin{equation}
D_{B}=
\begin{pmatrix}
0&0&0&{i} \\0&0&-{i}&0\\0&{i}&0&0 \\-{i}&0&0&0
\end{pmatrix}.
\end{equation}
The eigenvalues and the corresponding eigenstates are
\begin{equation}
\begin{aligned}
+1 \quad
|\tilde{0}\rangle_{B}&=\frac{1}{\sqrt{2}}(|00\rangle_{B}-{i}|11\rangle_{B}),\quad |\tilde{1}\rangle_{B}=\frac{1}{\sqrt{2}}(|01\rangle_{B}+{i}|10\rangle_{B});\nonumber\\
-1 \quad
|\tilde{0}\rangle_{B}&=\frac{1}{\sqrt{2}}(|01\rangle_{B}-{i}|10\rangle_{B}),\quad |\tilde{1}\rangle_{B}=\frac{1}{\sqrt{2}}(|00\rangle_{B}+{i}|11\rangle_{B}).
\end{aligned}
\end{equation}
Therefore, when we fix $D_{l}=1$ or $D_{l}=-1$, the four-dimensional Hilbert space of each site reduces to two dimensions expanded by the two corresponding eigenstates. The new $a$-fermion basis $|0_{a}\rangle_{D_l}, |1_{a}\rangle_{D_{l}}$ are just the eigenstates $|\tilde{0}\rangle _{l}, |\tilde{1}\rangle_{l}$ of $D_{l}$. It is straightforward to check the relation between the bases and corresponding operators using the Majorana basis in Eq. (\ref{mbasis}) and Dirac basis in Eq. (\ref{abasis}).
\begin{equation}
a_{l}|\tilde{0}\rangle_{l}=0,\quad a_{l}|\tilde{1}\rangle_{l}=|\tilde{0}\rangle_{l},\quad
a_{l}^{\dagger}|\tilde{1}\rangle_{l}=0,\quad a_{l}^{\dagger} |\tilde{0}\rangle_{l}=|\tilde{1}\rangle_{l},
\end{equation} where $l\in A,B$ .  What is more, the anticommutation relations of $a$ fermions can also be checked:
\begin{equation}
 \{a^{\dagger}_{i},a_{j} \}=\delta_{ij},\quad \{a_{i},a_{j} \}=0.
\end{equation}
The zero-energy states are just the fully occupied and vacuum states of
$a$ fermions,
\begin{equation}
\begin{aligned}
|s^{+}_{U}\rangle&=\prod_{i\in A,j\in B} |\tilde{0}\rangle_{i}|\tilde{0}\rangle_{j}=\prod_{l\in A,B}|0_{a}\rangle_{D_l=+1};
\nonumber\\
|s^{-}_{U}\rangle&=\prod_{i\in A,j\in B} |\tilde{1}\rangle_{i}|\tilde{1}\rangle_{j}=\prod_{l\in A,B}|1_{a}\rangle_{D_l=-1},
\end{aligned}
\end{equation}
and written in the $|n_{\uparrow},n_{\downarrow}\rangle$ basis
\begin{equation}
\begin{aligned}
|s^{+}_{U}\rangle=&\prod_{i\in A,j\in B}\frac{1}{\sqrt{2}}(|00\rangle_{i}+i|11\rangle_{i})\frac{1}{\sqrt{2}}(|00\rangle_{j}-i|11\rangle_{j});\\
|s^{-}_{U}\rangle=&\prod_{i\in A,j\in B}\frac{1}{\sqrt{2}}(|00\rangle_{i}-i|11\rangle_{i})\frac{1}{\sqrt{2}}(|00\rangle_{j}+i|11\rangle_{j}).
\end{aligned}
\end{equation}
Then, undo the unitary transformation $U=\prod_{i\in A, j\in B}{\rm exp}[{{i}\pi/2(\tilde{c_{i}}^{\dagger}\tilde{c_{i}}-\tilde{c_{j}}^{\dagger}\tilde{c_{j}})}]$ and map the states back to the density matrix:
\begin{equation}
\begin{aligned}
|s^{+}\rangle&=\prod_{l\in A,B}\frac{1}{\sqrt{2}}(|00\rangle_{l}-|11\rangle_{l})\to\rho^{+}=\frac{1}{2}(|0\rangle\langle0|-|1\rangle\langle1|);\\
|s^{-}\rangle&=\prod_{l\in A,B}\frac{1}{\sqrt{2}}(|00\rangle_{l}+|11\rangle_{l})\to\rho^{-}=\frac{1}{2}(|0\rangle\langle0|+|1\rangle\langle1|).
\end{aligned}
\end{equation}

\bibliography{solve}

\end{document}